\documentclass[final,5p,times]{elsarticle}
\usepackage{graphicx}% Include figure files
\usepackage{dcolumn}% Align table columns on decimal point
\usepackage{bm}% bold math
\usepackage{color}
\usepackage{soul}

\begin{document}
\begin{frontmatter}

\title{Phase space deformations in SUSY  cosmology.}

\author{J. L. L\'opez-Picón$^a$\corref{cor1}}
\ead{jl_lopez@fisica.ugto.mx}
\author{M. Sabido$^{a}$\corref{cor2}}
\ead{msabido@fisica.ugto.mx}
\author{C. Yee-Romero$^b$\corref{cor3}}
\ead{carlos.yee@uabc.edu.mx}

\address{  $^a$ Departamento  de F\'{\i}sica de la Universidad de Guanajuato,\\
 A.P. E-143, C.P. 37150, Le\'on, Guanajuato, M\'exico.\\
 $^b$ Departamento de Matem\'aticas, Facultad de Ciencias\\
 Universidad Aut\'onoma de Baja California, Ensenada, Baja California, M\'exico.\\
}%
\begin{abstract}
In this paper we propose  a SUSY generalization for deformed phase-space cosmology.  In particular, scalar field and phantom cosmology are studied.
We construct  the supercharges of the models and  the SUSY Wheeler-DeWitt equation. We also construct and   derive the classical dynamics using the WKB approximation.
\end{abstract}

\end{frontmatter}

%%%%%%%%%%%%%%%%%%%%%%%%%%%%%%%%%%%%%%%%%%%%%%%%
%%%%%%%%%%%%%%%%%%%%%%%%%%%%%%%%%%%%%%%%%%%%%%%%
\section{Introduction}
%%%%%%%%%%%%%%%%%%%%%%%%%%%%%%%%%%%%%%%%%%%%%%%%
%%%%%%%%%%%%%%%%%%%%%%%%%%%%%%%%%%%%%%%%%%%%%%%%
By uniting General Relativity (GR) with the standard model of particles, the $\Lambda CDM$ model has been established as the most accurate description of the Universe.  With some basic assumptions, it accounts for inflation, perturbations, structure formation and  the current acceleration of the Universe. Although the $\Lambda$CDM model is compatible with observations, as it is a  phenomenological model, it is not surprising that several theoretical aspects remain unsolved.   
For example, the growing tension between the Planck observations of the cosmic microwave background anisotropies and the local measurement of the Hubble constant \cite{tension,vanPutten:2017bqf,Colgain:2018wgk,vanPutten,Agrawal:2018own}. Or the unresolved dark energy problem, as proposing a cosmological constant entails inconsistencies with traditional quantum field theory \cite{weinberg,lambda}. In particular, scalar fields have been successful as an alternative
 for the description of dark energy \cite{SF1,ratra,SF2,SF3}. 
Also, phantom fields  (scalar fields  negative kinetic term) have been considered in the literature. These fields  provide an effective negative pressure and a repulsive effect that in the long term could be responsible for the late time accelerated expansion \cite{Caldwell}. 
This amounts to suggesting that our understanding of the gravitational interaction is complete, and simply need to introduce the appropriate matter to Einstein's equations.
Therefore, if  the dark energy problems are related to the poor understanding of gravity then  modifications to gravity should be considered.

Supersymmetry revolutionized the landscape of theoretical physics {\cite{wess_bagger}. This new symmetry}, allowed the transformation from bosons to fermions and vice versa. It was constructed in the context of particle physics, but eventually the local gauge theory of supersymmetry resulted in supergravity {\cite{nieuwenhuizen}}, the  supersymmetric (SUSY) generalization of general relativity (GR). {After the realization that supegravity is the square root of GR \cite{teitelboim1,teitelboim2}, cosmological models were  studied.
Following these ideas, supersymmetric quantum cosmology was proposed in \cite{octavio}. From the square root of the Wheeler-DeWitt (WDW) equation, SUSY generalization of cosmological models can be derived. One can construct the SUSY  Hamiltonian of the cosmological model, by defining the ``square root'' of the potential in the minisuperspace variables \cite{graham,lidsey}. Also, a superfield formulation was constructed. This approach allows to introduce matter the cosmological model \cite{tkach}. Research in the SUSY version of the WDW equation \cite{death}, led to several approaches to SUSY quantum cosmology (for complete and up to date  review o SUSY quantum cosmology, see \cite{moniz_1,moniz_2}).}\\ 
{At the beginning of the century, the old idea of a noncommutative space time was rekindled. This idea was exploited in particle physics, but eventually studied in the context of gravity. Several noncommutative versions of gravity where constructed \cite{Obregon1,Calmet,Wess,NC2}. As noncommutative effects are expected to be present near Planck's scale, one can  consider an inherently noncommutative spacetime at the early ages of the universe. Unfortunately, using noncommutative  gravity is a complicated ordeal,  due to the highly nonlinear character of these theories. Following the lessons from SUSY cosmology, in \cite{Obregon2} the authors introduce} the effects of noncommutativity  using the methods of nocommutative quantum mechanics \cite{gamboa} on the WDW equation to construct noncommutative quantum cosmology. Moreover,  classical effects of the  noncommutative deformation where explored  using the WKB approximation of the noncommutative quantum model \cite{eri1,vakili, Miguel1,yee,eri2,Miguel2,huicho,shiraishi,rasouli,rasouli2,rasouli3,rasouli4,saba}. 
  
The main objective of this paper is to construct a consistent model for SUSY noncommutative quantum cosmology and analyze the classical limit  of noncommutative SUSY cosmology.
We will consider two models,  a Friedmann-Robertson-Walker (FRW) cosmological model coupled to a scalar field and an one coupled to a phantom  field. We will introduce SUSY as well as a phase space deformation in the minisuperspace, to derive the SUSY noncommutative WDW equation. We study the  quantum models and using the semiclassical approximation of the SUSY deformed quantum model {we outline} the classical implications.% 

The paper is organized as follows. In section 2, we review the bosonic model and solve the deformed WDW equation. The SUSY generalization is done in section 3, we find the  deformed SUSY WDW equation 
and fix the parameters in the deformation in order to make the quantization possible. We also show that the classical paths  arise from the semiclassical approximation of the SUSY WDW equation obtained from the deformed Hamiltonian. Finally, section 4 is devoted for concluding remarks.   
%%%%%%%%%%%%%%%%%%%%%%%%%%%%%%%%%%%%%%%%%%%%%%%%         
%%%%%%%%%%%%%%%%%%%%%%%%%%%%%%%%%%%%%%%%%%
\section{The Bosonic Model}
%%%%%%%%%%%%%%%%%%%%%%%%%%%%%%%%%%%%%%%%%%
%%%%%%%%%%%%%%%%%%%%%%%%%%%%%%%%%%%%%%%%%%%%%%%%
In this section we review the deformed phase space bosonic model, for both the scalar field \cite{Miguel1} and the phantom field \cite{huicho}.

Let us start  with  the flat FRW metric
\begin{equation}
ds^2 = -N^2(t)dt^2 + a^2(t)[ dr^2 + r^2d\Omega^2],
\end{equation}
where $a(t)$ and $N(t)$ correspond to the scale factor and the lapse function respectively. Over this background, the action of a minimally coupled  scalar field $\varphi (t)$ with constant potential will be   
\begin{equation}\label{accion}
S = \int dt \left \{\   -\frac{3a \dot{a}^2}{N} - a^3\left(  \epsilon\frac{\dot{\varphi}^2}{2N} + N\Lambda    \right)    \right \}, \label{action}
\end{equation}   
setting the units as $8\pi G = 1$. We have introduced deliberately the factor $\epsilon$ in the kinetic part of the scalar action in order to include the usual scalar field and a phantom scalar field, therefore 
$\epsilon$ can take the values $\epsilon = 1$ and $\epsilon = -1$ respectively. 
The canonical Hamiltonian derived from Eq.(\ref{accion}) is
\begin{equation}\label{hamcon}
H=-N \left[ \frac{P_a^2}{12a} +\epsilon \frac{P_{\varphi}^2}{2a^3} - a^3 \Lambda \right]. \label{Ham}
\end{equation}
For the usual scalar field $\epsilon=1$, we make the change of variables 
$ x = m^{-1} a^{3/2} \sinh ( m \phi )$, $y = m^{-1} a^{3/2} \cosh (m \phi)$.
For the phantom scalar field $\epsilon=-1$, we use a different transformation 
$x = \mu^{-1}a^{3/2}\sin{(\mu \varphi)}$, $y = \mu^{-1}a^{3/2}\cos{(\mu \varphi)},$
with  $\mu = \sqrt{3/8}$ and the Hamiltonian Eq.(\ref{hamcon}) can be rewritten as 
\begin{equation}\label{hamiltonian}
H = N\left(  \frac{1}{2}P_x^2 + \frac{\omega^2}{2}x^2   \right) + \epsilon N\left(  \frac{1}{2}P_y^2 + \frac{\omega^2}{2}y^2   \right), \label{Ham1}
\end{equation}
where $\omega^2 = -\frac{3}{4}\Lambda$. 
When one considers the usual scalar field, the Hamiltonian is transformed to a ``ghost oscillator'' which is simply a difference of two harmonic oscillators  and for the phantom scalar field, the Hamiltonian is the sum of two harmonic oscillators.

Several approaches have been considered to incorporate noncommutativity into physical theories and 
cosmology is one of many examples. There is a broadly explored path with the aim of studying noncommutativity \cite{Obregon2}, where the noncommutative deformation is performed in the  minisuperspace variables. The procedure we will follow is based on this approach, but the deformation is done in the minisupespace variables and their associated momentum \cite{Miguel1}. We start with  the  transformation
 \begin{eqnarray}\label{nctrans}
\widehat{x}&=&x+\frac{\theta}{2}P_{y}, \quad \widehat{y}=y-\frac{\theta}{2}P_{x},\\
\widehat{P}_{x}&=&P_{x}-\frac{\beta}{2}y, \quad \widehat{P}_{y}=P_{y}+\frac{\beta}{2}x, \nonumber
\end{eqnarray}
on the classical phase space variables $\{x,y,P_x,P_y\}$. These are the variables that satisfy the usual Poisson algebra and the new variables satisfy a deformed algebra
\begin{equation}\label{algebra}
\{\widehat{y},\widehat{x}\}=\theta,\; \{\widehat{x},\widehat{P}_{x}\}=\{\widehat{y},\widehat{P}_{y}\}=1+\sigma,\; \{\widehat{P}_y,\widehat{P}_x\}=\beta,\label{dpa}
\end{equation}
where $\sigma=\theta\beta/4$. {It is important to mention, that the deformation in Eq.(\ref{algebra}) is a particular choice. Alternative relations for the algebra, have been explored  to analyze different physical scenarios (i.e. gravitational collapse \cite{rasouli}, inflation \cite{rasouli2}, chameleon dynamic's \cite{saba}, among others).}

In order to establish the deformed theory, the starting point is the analogous Hamiltonian to Eq.(\ref{Ham1}) 
but constructed with the variables obeying the modified algebra Eq.(\ref{dpa}). For the usual scalar field the deformed Hamiltonian is
\begin{equation}
H_{nc} = \frac{1}{2} \left[   (P_x^2 - P_y^2) + \ell_s^2(xP_y+yP_x) + {\omega}_s^2(x^2-y^2)    \right], \label{Ham2_scalar}
\end{equation}
where $\ell_s^2$ and ${\omega_s}^2$ are
\begin{equation}
\ell_s^2 = \frac{\beta - \omega^2 \theta}{1 - \frac{\omega^2 \theta^2}{4}}, ~~~~~{\omega_s}^2 = \frac{\omega^2 - \frac{\beta^2}{4}}{1 - \frac{\omega^2 \theta^2}{4}}.
\end{equation}
For the phantom model
\begin{equation}
H_{nc}=\frac{1}{2} \left[ (P_x^2 + P_y^2) + \ell_p^2(xP_y-yP_x) + {\omega_p}^2(x^2+y^2) \right], \label{Ham2}
\end{equation}
where $\ell_p^2$ and ${\omega_p}^2$ are
\begin{equation}
\ell_p^2 = \frac{\beta + \omega^2 \theta}{1 + \frac{\omega^2 \theta^2}{4}}, ~~~~~{\omega_p}^2 = \frac{\omega^2 + \frac{\beta^2}{4}}{1 + \frac{\omega^2 \theta^2}{4}}.
\end{equation}
Important differences between the transformed Hamiltonian\footnote{{The deformed phase space scalar field model was studied in \cite{Miguel1}, were a relationship between the cosmological constant and the deformation parameters was established.}} of the scalar field cosmology model \cite{Miguel1} and  the deformed Hamiltonian of the phantom cosmology model Eq.(\ref{Ham2}) should be noticed. 
The crossed term involving products of position and momentum in the phantom model corresponds to an 
angular momentum term, which is not the case in the scalar field Hamiltonian. 
The deformation Eq.(\ref{nctrans}) represents two nonequivalent physical descriptions, the {\it``C-frame"}
and the {\it``NC-frame"}. The effects of the deformation are interpreted as a commutative 
space with a modified interaction in the first, and where we work with the deformed variables and the original interaction in the second.  
For the rest of the paper we will work in   
the {\it``C-frame"} as the interpretation is clearer. For example, we can interpret the deformed model as a bidimensional harmonic oscillator  
and the minisuperspace deformation comes into play as an angular momentum term in the Hamiltonian.  

For the phantom field, the quantization of the deformed Hamiltonian  is {straightforward}%(unlike the regular scalar field). 
Using the canonical quantization approach $H_{nc} \psi = 0$, gives the corresponding WDW equation. 
The Hamiltonian $H_{nc}$, explicitly contains the angular momentum operator $(xP_y - yP_x) = L_z$
and can be written in terms of the Hamiltonian of a 2-dimensional harmonic oscillator  $H_{xy}$, with frequency ${\omega_p}^2$ plus an angular momentum term, that is $H_{nc} = H_{xy} + L_z$. Using the similarity of the WDW equation, with the 2-dimensional harmonic oscillator, we obtain 
the wave function  of the Universe $\psi_{n,m} (r,\phi)$  

\begin{equation}\label{Wave}
\psi_{n,m}(r,\phi) = N \left( \frac{{\omega_p}r^2}{\hbar} \right)^{\frac{|m|}{2}} L^{|m|}_n\left( \frac{{\omega_p}r^2}{\hbar} \right)e^{im\phi}e^{-\frac{\widetilde{\omega_p} r^2}{2 \hbar}}. \end{equation}
where  $n=0,1,2,...\:; ~m = 0, \pm 1, \pm 2,...,\pm n,$ and $L_n^{\alpha} (z)$  are the generalized Laguerre polynomials. 
The zero energy condition 
$E = 0$ from the WDW equation  is trivially satisfied by $\ell = {\omega_p} = 0$. For 
$\ell \neq 0, ~{\omega_p} \neq 0$ (for instance for a single state of quantum numbers $n$ and $m$) we get the general equation  
$
{ m\ell^2} = -{2(2n + |m| + 1)}{{\omega_p}}. 
$ 

Finally, one can apply a WKB type approximation  on the WDW equation $H_{nc} \Psi = 0$, to verify that we get the same classical dynamics.  The two resulting equations are the classical equations of motion. Therefore we can be confident that  
the same classical paths arise in the classical limit of the  WDW equation obtained from the deformed Hamiltonian.   

%%%%%%%%%%%%%%%%%%%%%%%%%%%%%%%%%%%%%%%%%%%%%%%%
%%%%%%%%%%%%%%%%%%%%%%%%%%%%%%%%%%%%%%%%%%%%%%%%
\section{The SUSY Model.}
%%%%%%%%%%%%%%%%%%%%%%%%%%%%%%%%%%%%%%%%%%%%%%%%
%%%%%%%%%%%%%%%%%%%%%%%%%%%%%%%%%%%%%%%%%%%%%%%%

Several approaches have been suggested to supersymmetrize the WDW-equation for cosmological models. 
The first {models\} proposed \cite{octavio}, {emerged} shortly after the appearance of supergravity. They show, that this theory provides a natural  square root of gravity. 
We follow an alternative method that allows to define a “square root” of the potential, in the minisuperspace, 
of the cosmological model of interest and consequently it is possible to find operators which 
square results in the Hamiltonian 
\cite{lidsey, moniz_1}. One can summarize this approach as an application of the methods of SUSY quantum mechanics 
to quantum cosmology. 
In this approach the Hamiltonian can be written as
\begin{equation}
    2H_0 = G^{\mu \nu}\Pi_{\mu}\Pi_{\nu} + U(\bar{q}),
\end{equation}
where $G^{\mu \nu}$ represents the metric in the minisuperspace and the classical potential $U(\bar{q})$ 
with $\bar{q}= (x,y)$ is related to the superpotential $\phi$ in the following way
\begin{equation}
    G^{\mu \nu} \frac{\partial \phi}{\partial q^{\mu}}\frac{\partial \phi}{\partial q^{\mu}} = U(\bar{q}).
\end{equation}
The minisuperspace Hamiltonian $H$ becomes 
\begin{equation}
    H = \frac{a}{2}\left(Q\bar{Q} + \bar{Q}Q \right) = H_0 
    + \frac{\partial^2 \phi}{\partial q^{\mu} \partial q^{\nu}}[\bar{\Theta}^{\mu},\Theta^{\nu}]
\end{equation}
where $Q$ and $\bar{Q}$ are the supercharges. The supercharges satisfy the closed superalgebra
\begin{eqnarray}
&&\left \{Q,\bar{Q}\right\}=2H,\\
&&\{Q,Q\}= \{\bar{Q},\bar{Q}\} =  0,[Q,H] = [\bar{Q},H]=0.\nonumber
\end{eqnarray}
To construct the supercharges, we  follow the formulation given in \cite{julio},
where the supercharges are given by
\begin{equation}
Q=\Theta^\mu\left( \Pi_\mu+i\frac{\partial\phi}{\partial q^\mu} \right), \quad \bar{Q}=\bar{\Theta}^\mu\left( \Pi_\mu-i\frac{\partial\phi}{\partial q^\mu} \right),
\end{equation}
where $\Pi^{\mu}$ are the bosonic momenta. The fermionic variables $\theta^{\mu}$, $\bar{\theta}^{\nu}$ satisfy the spinor algebra
\begin{equation}
\{\bar{\Theta}^\mu,\bar{\Theta}^\nu\}=0,\quad\{\Theta^\mu,\Theta^\nu\}=0,\quad \{\bar{\Theta},\Theta\}=G^{\mu\nu}.
\end{equation}
For the scalar field, $G^{\mu\nu}=diag(1,-1)$ with $U(x,y)=1/2(x^2-y^2)$ and for the phantom field $G^{\mu\nu}=diag(1,1)$ with $U(x,y)=1/2(x^2+y^2)$.

A particular representation for fermionic variables $\Theta^{\mu}$, $\bar{\Theta}^{\nu}$ can be constructed for the scalar and phantom fields or equivalently for the difference and sum of harmonic oscillators.
For the phantom field, we have
\begin{eqnarray}
  \Theta^{x}_p = \frac{1}{2\sqrt{2}}\left( i\gamma^1 + \gamma^2 \right),~~~\bar{\Theta}^{x}_p = \frac{1}{2\sqrt{2}}\left( i\gamma^1 -  \gamma^2 \right),\\
  \Theta^{y}_p = \frac{1}{2\sqrt{2}}\left( \gamma^0 + \gamma^3 \right), ~~~\bar{\Theta}^{y}_p = \frac{1}{2\sqrt{2}}\left( \gamma^0 -  \gamma^3 \right),~\ \nonumber
\end{eqnarray}
\noindent where $\gamma^{\mu}$ are the Gamma matrices
\begin{equation}
\hspace{-.5cm}
\gamma^{0} = \left(
 \begin{array}{cccc}
 0 & 0 & 0 & -i \\
 0 & 0 & i & 0 \\
 0 & -i & 0 & 0 \\
 i & 0 & 0 & 0 
 \end{array} \right),~~\gamma^{1} = \left(
 \begin{array}{cccc}
 i & 0 & 0 & 0 \\
 0 & -i & 0 & 0 \\
 0 & 0 & i & 0 \\
 0 & 0 & 0 & -i 
 \end{array} \right), 
\end{equation} 
\[
\hspace{-.5cm}
 \gamma^{2} = \left(
 \begin{array}{cccc}
 0 & 0 & 0 & i \\
 0 & 0 & -i & 0 \\
 0 & -i & 0 & 0 \\
 i & 0 & 0 & 0 
 \end{array} \right),~~\gamma^{3} = \left(
 \begin{array}{cccc}
 0 & -i & 0 & 0 \\
 -i & 0 & 0 & 0 \\
 0 & 0 & 0 & -i \\
 0 & 0 & -i & 0 
 \end{array} \right). 
\]

\noindent For the scalar field
\begin{eqnarray}
  \Theta^{x}_s = \frac{1}{2\sqrt{2}}\left( \gamma^1 -i \gamma^2 \right),~~~\bar{\Theta}^{x}_s = \frac{1}{2\sqrt{2}}\left( \gamma^1 +  i\gamma^2 \right), \\
  \Theta^{y}_s = \frac{1}{2\sqrt{2}}\left( \gamma^0 + \gamma^3 \right), ~~~\bar{\Theta}^{y}_s = \frac{1}{2\sqrt{2}}\left( \gamma^0 -  \gamma^3 \right),~\ \ \nonumber
\end{eqnarray}
\noindent where $\gamma^{\mu}$ are the same  Gamma matrices. When one considers a usual scalar field, the Hamiltonian associated to the action Eq.(\ref{action}) corresponds to a ghost oscillator, 
namely the difference of two harmonic oscillators.

The original construction of noncommutative quantum cosmology \cite{Obregon2}, was based on applying the ideas of noncommutative quantum mechanics \cite{gamboa} to the WDW equation of the Kantowski Sachs cosmological model. Therefore, it makes sense to follow this line of reasoning and apply the approach of non commutative supersymmetric quantum mechanics \cite{das}, to quantum cosmology.
%Comparing  the deformed Hamiltonian with \textcolor{blue}{a} commutative model immersed in a constant perpendicular magnetic field $B$, we conclude that 
{Moreover, the effects of the minisuperspace deformation can be interpreted as a commutative theory with a conserved charge.} 
Consequently for the deformed SUSY model, we will follow the usual SUSY quantum mechanics and introduce the full deformation 
(coordinates and momenta) as minimal coupling. This gives the same deformed Hamiltonian Eq.(\ref{Ham2_scalar}) that we 
get from using the deformed algebra Eq.(\ref{dpa}), but with a different frequency for the potential.

Let us begin with the {phantom field} where as already mentioned, the Hamiltonian is a bidimensional harmonic oscillator. 
From  the potential of Eq.(\ref{hamiltonian}) we calculate the superpotential $\phi$, then the resulting supercharges are
\begin{eqnarray}
 Q = \Theta^x_p\left( P_x-\frac{\omega^2_p}{2}y+ i\ell_p x\right) + \Theta^y_p\left( P_y+\frac{\omega^2_p}{2}x+ i\ell_p y\right),\\ 
 \bar{Q}= \bar{\Theta}^x_p\left( P_x-\frac{\omega^2_p}{2}y- i\ell_p x\right) + \bar{\Theta}^y_p\left( P_y+\frac{\omega^2_p}{2}x- i\ell_p y\right),\nonumber
\end{eqnarray}
where we have used the vector potential ${A}_{x}=\frac{\omega_1^2}{2}{y}$ and ${A}_{y}=-\frac{\omega_1^2}{2}{x}$.

Using the SUSY algebra we get the Hamiltonian operator (after diagonalization)
\begin{eqnarray}\label{susy_phantom}
&&H= \frac{1}{2}\left [P_x^2 + P_y^2 + \ell_p^2\left(xP_y-yP_x\right) +\right.\\
&&\left .\left( \ell_p^2+\frac{\omega_p^4}{4}\right)\left( x^2+y^2\right) \right]\mathbb{I}
+\frac{\ell_p^2}{2}\mathbb{A}^++\omega_p \mathbb{B}^+,\nonumber
\end{eqnarray} 
where $\mathbb{A}^+=diag(-2,0,0,2)$ and $\mathbb{B}^+=diag(0,2,-2,0)$.

We followed the same approach for the scalar field and the resulting supercharges are
\begin{eqnarray}
 Q = \Theta^x_s\left( P_x-\frac{\omega^2_s}{2}y+ i\ell_s x\right) + \Theta^y_s\left( P_y+\frac{\omega^2_s}{2}x+ i\ell_s y\right),\\ 
 \bar{Q}= \bar{\Theta}^x_s\left( P_x-\frac{\omega^2_s}{2}y- i\ell_s x\right) + \bar{\Theta}^y_s\left( P_y+\frac{\omega^2_s}{2}x- i\ell_s y\right),\nonumber
\end{eqnarray}
as in the phantom case, we use the vector potential ${A}_{x}=\frac{\omega_1^2}{2}{y}$ and ${A}_{y}=-\frac{\omega_1^2}{2}{x}$. The Hamiltonian operator (after diagonalization) {becomes}
\begin{eqnarray}
&&H= \frac{1}{2}\left [P_x^2 - P_y^2 + \ell_s^2\left(xP_y+yP_x\right) +\right.\\
&&\left .\left( \frac{\omega_s^4}{4}-\ell_s^2\right)\left( x^2-y^2\right) \right]\mathbb{I}
+\frac{\omega_s^2}{2}\mathbb{A}^-+\ell_s \mathbb{B}^-,\nonumber
\end{eqnarray} 
where $\mathbb{A}^-=diag(0,0,-2i,2i)$ and $\mathbb{B}^-=diag(-2,2,0,0)$.

To obtain the  SUSY WDW equation, we proceed as in \cite{julio_octavio} and use canonical quantization on the deformed Hamiltonian. Then, from the usual canonical quantization approach $H_{nc} \psi = 0$, we can obtain the corresponding SUSY WDW equation. 
Let us consider the Hamiltonian in Eq.(\ref{susy_phantom})
The Hamiltonian $H_{nc}$, explicitly contains the angular momentum operator 
and can be written in terms of the Hamiltonian of a 2-dimensional harmonic oscillator $H_{xy}$, with frequency $\widetilde{\omega}^2$ plus an angular momentum term, that is 
$H_{nc} = H_{xy} + L_z$. Using the similarity of the WDW equation with the 2-dimensional harmonic oscillator, 
first we write the eigenvalue equation for the Hamiltonian, then we will find the energy eigenvalues and impose the zero energy condition. In  polar coordinates the NC-SUSY WDW equation is given by

\begin{equation}
\hspace{-0.5cm}
-\frac{\hbar^2}{2}\left[  \frac{\partial^2 \psi}{\partial r^2} + \frac{1}{r}\frac{\partial \psi}{\partial r} + \frac{1}{r^2}\frac{\partial^2 \psi}{\partial \phi^2}  \right] 
+ \frac{\widetilde{\omega}^2}{2}r^2\psi -\frac{i\hbar \ell^2}{2} \frac{\partial \psi}{\partial \phi} = C \psi.~~ \label{WDW}
\end{equation} 
The wave function can be written in terms of the eigenstates of the Hamiltonian and angular momentum since both operators commute. 
Every state is uniquely specified by the quantum numbers $n$ and $m$. The wave function solution to Eq.(\ref{WDW}) is  the same as in the bosonic case, and is given by Eq.(\ref{Wave}),

%\begin{equation}\label{Wave}
%\psi_{n,m}(r,\phi) = N \left( \frac{\widetilde{\omega}r^2}{\hbar} \right)^{\frac{|m|}{2}} %L^{|m|}_n\left( \frac{\widetilde{\omega}r^2}{\hbar} \right)e^{im\phi}e^{-\frac{\widetilde{\omega} %r^2}{2 \hbar}}. \end{equation}
%where  $n=0,1,2,...\:; ~m = 0, \pm 1, \pm 2,...,\pm n,$ and $L_n^{\alpha} (z)$  are the generalized Laguerre polynomials. 
Unlike the bosonic case, we do not need to satisfy the zero energy condition, as we can consider
$E = C$. For 
$\ell \neq 0, ~\widetilde{\omega} \neq 0$ we get the general equation  $
{m}{\ell^2} +2{\widetilde{\omega}}(2n + |m| + 1)=m {\widetilde{\omega}}C.$ By using the definitions for $\ell^2$ and $\widetilde{\omega}^2$ we can write this equation as
\begin{equation}
\frac{16\beta -12\theta \Lambda}{\sqrt{(16-3\theta^2 \Lambda)(4\beta^2-12\Lambda)}}  +\frac{2(2n + |m| + 1)}{m}=C. \label{Condition3}
 \end{equation}
With respect to the requirement in Eq.(\ref{Condition3}), we need to find the values of $\theta$, $\beta$ and $\Lambda$ to satisfy the condition. 
Now we 
write the wave function %$\psi_{n,m} (a,\varphi)$ 
in terms of the scale factor $a$ and the  field $\varphi$ to construct a general wave packet 
\begin{equation}
 \psi(a,\varphi) = \sum_n^N c_{n,m}\psi_{(n,m)} (a,\varphi), \label{WaveP}  
\end{equation}  
 with $m$ fixed and $c_{n,m}$ as arbitrary constants as long as the condition $E = \sum_n^N E_{(n,m)} = C$  holds. 
 This allows to find values for $\theta$, $\beta$ and $\Lambda$ 
that satisfy the condition Eq.(\ref{Condition3}).

For the scalar field, the Hamiltonian is the difference of two harmonic oscillators and the quantization of this model is not  {straightforward} as in the previous case. Moreover, the extra term that appears in the deformed Hamiltonian is not an angular momentum term.\\
%For a given value of $m$, the sum in Eq.(\ref{WaveP}) runs over odd/even values of $n$
%depending on the odd/even value of $m$ respectively. The wave packet is normalizable by the use of 
%the orthogonality relation of generalized Laguerre polynomials.
%The total energy of the wave packet %Eq.(\ref{WaveP}) is given explicitly by%
%
%\begin{equation}
%E = \hbar \widetilde{\omega}N^2 + 2|m|\hbar \widetilde{\omega} N + mN\hbar \frac{\ell^2}{2}=C,
%\end{equation}
%will satisfy the condition in Eq.(\ref{Condition3}).
%From  the arguments given above, we conclude that it is always possible to find, for a given $\Lambda$,  values for $\theta$ and $\beta$ for which the condition holds. 
To find the classical solutions we follow the ideas in \cite{julio_octavio,julio_octavio2}. We simply take  the WKB approximation on the noncommutative SUSY WDW equation. After diagonalization of the Hamiltonian and acting on the wave function $\Psi = (\psi_1,\psi_2,\psi_3,\psi_4)$, one gets four equations.
Starting with the phantom case, the equations are
\begin{eqnarray}\label {4H}
(H_{NC} -\ell_p^2)\psi_1 &=& 0,\quad (H_{NC} +2\omega_p)\psi_2 = 0,\\ 
(H_{NC} -2\omega_p)\psi_3 &=& 0,\quad 
(H_{NC} +\ell_p^2)\psi_4 = 0.\nonumber 
\end{eqnarray}
Now we perform the semiclassical WKB approximation for a wave function of the form $\psi = \exp{\frac{i}{\hbar} S_{1}(x) + \frac{i}{\hbar} S_{2}(y)}$. Using the approximation $ \left( \frac{\partial S_1(x)}{\partial x} \right)^2 > > \frac{\partial^2 S_1(x)}{\partial x^2}$ and $ \left( \frac{\partial S_2(y)}{\partial y} \right)^2 > > \frac{\partial^2 S_2(y)}{\partial y^2}$  on Eq.(\ref{4H}), we get the corresponding Hamilton-Jacobi equations
\begin{eqnarray}
&&\left( \frac{\partial S_1(x)}{\partial x} \right)^2 + \left( \frac{\partial S_2(y)}{\partial y} \right)^2 + {\omega}_p^2(x^2 + y^2)\\
&& + \ell_p^2 \left( \frac{\partial S_2(y)}{\partial y} \right)x - \ell_p^2 \left( \frac{\partial S_1(x)}{\partial x} \right)y \pm 2C = 0,\nonumber
\end{eqnarray}
because the only difference between the four equation is the constant, $C$ takes the values $\ell_p$ or $\omega_p$ to reproduce the four equations. By identifying  $\frac{\partial S_1(x)}{\partial x} = p_x$  and $\frac{\partial S_2(y)}{\partial y} = p_y$  we get
\begin{equation}
\dot{x}^2 + \dot{y}^2 + \widetilde{\omega}_p^2 (x^2 + y^2) \pm 2C = 0, 
\end{equation}
where we  have defined $\widetilde{\omega}_p=\omega_p^2-\frac{\ell_p^4}{4}$. Moreover, 
differentiating the equation and dividing by $\dot{x}\dot{y}$ we have 
\begin{equation}
\frac{\ddot{x}}{\dot{y}} + \widetilde{\omega}_p^2\frac{x}{\dot{y}} + \frac{\ddot{y}}{\dot{x}} + \widetilde{\omega}_p^2\frac{y}{\dot{x}} = 0, 
\end{equation} 
from which the classical equations of motion are recovered. Derivations 
allow for the supersymmetric contribution to disappear. These constants only have a real effect on the supersymmetric wave function solutions. The Hamiltonian takes the form
\begin{equation}
 \hspace{-.5cm}
 2H=\pi_x^2 + \pi_y^2 + \ell_p^2\left(x\pi_y-y\pi_x\right)
 +\widetilde{\omega}_p^2\left( x^2+y^2\right), \nonumber
\end{equation}
and the equations of motion are
\begin{equation}\label{EqMot}
\hspace{-.5cm}
 \ddot{x}+\ell^2 \dot{y} + \widetilde{\omega}_p^2x =0,\ \
 \ddot{y} -\ell^2 \dot{x} + \widetilde{\omega}_p^2y = 0,
\end{equation} 
these equations can be easily solved with the transformation $z=x+iy$.  We  have three different solutions depending on the sign of $\widetilde{\omega}_p^2$.

For the scalar field we follow the same approach. Starting from the NC-SUSY Hamiltonian, we can see that the four Hamiltonians in Eq.(\ref{4H}) have the same form
\begin{eqnarray}
 H &=& H_{NC} \pm C \\
 &=& p_x^2 + p_y^2 + {\omega}_s^2(x^2 + y^2) + \ell_s^2(xp_y + yp_x)  \pm C,\nonumber
\end{eqnarray} 
the we can write it as one equation, and the different hamiltonians correspond to different values of $C$. 
Using the WKB approximation, the classical limit the Hamiltonian's is
\begin{equation}
\hspace{-.5cm}
 2H=
\pi_x^2 - \pi_y^2 + \ell_s^2\left(x\pi_y+y\pi_x\right)
 +\widetilde{\omega}_s^2\left( x^2-y^2\right),
\end{equation}
and as in the previous case because the difference is only a constant, there are no contributions for the classical evolution.

\section{Conclusions}
In this paper we have constructed Noncommutative SUSY  cosmology.  
The noncommutative deformation was applied on the minisuperspace variables and  the construction was performed in the  {\it``C-frame"} where the deformation has a simple physical interpretation. %\footnote{\ms{The deformation is related to the introduction to an effective perpendicular and constant ``magnetic field" $\vec{B}$.}} 
%This can be seen by taking the bidimensional harmonic oscillator coupled to the vector potential $\vec A=(\frac{\ell^2}{2}y,-\frac{\ell^2}{2} x)$. 

The supersymmetric construction is {straightforward}, using minimal coupling on the supercharges form which SUSY deformed Hamiltonian. 
Using canonical quantization we obtain the  noncommutative SUSY WDW equation, and solving SUSY WDW equation  found the wave function of the Universe. 
Finally, we performed the WKB approximation  
and show that the same classical solutions that arise in the classical limit of the WDW equation of the corresponding 
deformed Hamiltonian are the same as in the bosonic case. Therefore, for this particular model, the inclusion of supersymmetry only introduces modification in the quantum limit by changing the zero energy condition. {This constant will be reflected in the solution spectrum by modifying the ground state.}

{The use of deformed-phase space has been used successfully to study several aspects of cosmology (i.e, dark energy, inflation, gravitational collapse). It warrants it is use, to tackle some of the remaining problems in SUSY quantum cosmology. The liberties allowed, when defining the deformed algebra can shed some insight to this problems. Because the deformation can be introduced as minimal coupling in the supercharges, deformed phase-space SUSY quantum cosmology is simple to implement. Furthermore, considering the different approaches \cite{moniz_1,moniz_2} to SUSY quantum cosmology, (in particular the  differential representation) more general models can be studied. This is an exciting perspective and will be reported elsewhere.}
    
\section*{Acknowledgments}
  
\noindent The authors would like to thank the anonymous referee who provided valuable comments and insight which helped to improve the manuscript. 
M. S. will like to thank the hospitality of the mathematics group (UABC-CA-44) of the Facultad de Ciencias. M. S. is supported by  CIIC 032/2023 and CIIC 224/2023.

\end{document}